\documentclass[aps,showpacs,a4paper,pre,twocolumn]{revtex4}
\usepackage{graphicx}

\usepackage{amsmath}
\usepackage{amssymb}

\newcommand{\PB}{\mathrm{PB}}
\newcommand{\ren}{\mathrm{ren}}
\renewcommand{\r}{\mathbf{r}}

\begin{document}

\title{ The polydisperse cell model: \\ Non-linear screening and charge renormalization in colloidal mixtures}
\date{\today}

\author{Aldemar Torres$^1$, Gabriel T\'ellez$^2$, Ren\'e van Roij$^1$}

\affiliation{$^1$ Institute for Theoretical Physics, Utrecht University \\
Leuvenlaan 4, 3584 CE Utrecht, The Netherlands.}

\affiliation{$^2$ Departamento de F\'\i sica, Universidad de Los Andes\\
A.~A.~4976 Bogot\'a, Colombia.  }

\begin{abstract}
We propose a model for the calculation of renormalized charges and
osmotic properties of mixtures of highly charged colloidal
particles. The model is a generalization of the cell model and the
notion of charge renormalization as introduced by Alexander and
his
 collaborators (\textit{J. Chem. Phys.} \textbf{80}, 5776 (1984)). The total solution is partitioned into
 as many different cells as components in the mixture. The radii of these cells are
determined self-consistently for a given set of parameters from
the solution of the non-linear Poisson-Boltzmann equation with
appropriate boundary conditions. This generalizes Alexanders's model
where the (unique) Wigner-Seitz cell radius is fixed solely by the colloids
packing fraction. We illustrate the technique by considering a
binary mixture of colloids with the same sign of charge. The present model can be used to calculate
thermodynamic properties of highly charged colloidal mixtures at
the level of linear theories, while taking the effect of non-linear
screening into account.
\end{abstract}

\pacs{82.70.Dd, 64.60.Cn, 64.10.+h, 82.39.Wj}

\keywords{Effective colloidal interactions, charge
renormalization, cell model.}

\maketitle

\section{Introduction}

Typical charge-stabilized colloidal suspensions consist of mesoscopic particles and ions, immersed in a solvent which to a good approximation can
be considered as a continuum medium, characterized by a dielectric permittivity $\epsilon$. The microscopic ions form a diffuse layer around the
colloidal surfaces, with a typical thickness of the order of the Debye length of the electrolyte $\kappa^{-1}$, and with a net charge that
compensates the charge of the colloidal surface. As a consequence, the effective colloidal interactions are screened: instead of the bare Coulomb
interaction $Z^2e^2/\epsilon{r}$ between two colloidal spheres of radius $a$ and charges $Ze$ a distance $r$ apart in the solvent, standard linear
Poisson-Boltzmann theory predicts a screened coulomb interaction $Z^2e^2\exp(2\kappa a)(1+\kappa a)^{-2}\exp[-\kappa{r}]/\epsilon{r}$ between the
colloids \cite{dlvo,Israel}. However, it is well-known that this relatively simple picture needs modification in the case of highly charged
colloids, where the strong electrostatic coupling between the colloids and their counterions induces the accumulation of the latter close to the
colloid-surface. This condensation phenomenon cannot be described by a linearized Poisson-Boltzmann approach, and requires the full nonlinear
Poisson-Boltzmann theory for a proper description \cite{Alexander}. However, the simplicity of the linear theories can be retained by the use of
the concept of charge renormalization, which considers each colloid and its condensed counterion shell as a single entity, carrying a net
effective charge which is usually smaller than the bare (structural) charge $Z$ of the colloid. This so-called renormalized charge, $Z^{\ren}$,
characterizes the long-distance behavior of the screened electric potential created by the colloid and its condensed microion polarization cloud,
and the effective colloidal pair interaction takes the linear-screening form but with $Z$ replaced by $Z^\ren$.

The concept of charge renormalization, and also its relatives such as dressed-ion theory \cite{Roland}, is very well-developed by now for
\emph{monodisperse} colloidal suspensions \cite{Alexander,Trizac,Trizac1,Trizac2}. The traditional theory by Alexander {\em et al.} is based on a
cell model, in which a suspension of $N$ colloidal spheres in an electrolyte of volume $V$ is approximated by $N$ identical spherical cells of
volume $V/N$, each containing a single colloid in the center and co- and counterions such that each cell is neutral. The simple spherically
symmetric geometry allows for a straightforward numerical solution of the nonlinear Poisson-Boltzmann equation in the cell, from which $Z^\ren$
follows directly \cite{Alexander,Trizac,Trizac1,Trizac2}. Although such an approach based on cells is mostly justified for colloidal crystals, the
concept of charge renormalization has proved to be very useful for fluid-state suspensions, e.g. in describing the structure factor
\cite{structurefactor} or the electrophoretic mobility \cite{schope}.

Much less is known about charge renormalization in the case of
colloidal \emph{mixtures} or in \emph{polydisperse} colloidal
suspensions. In recent years such mixtures of charged colloids
have generated a lot of interest, for instance because of their
ability to form a plethora of crystal phases \cite{crystals,phase}
or fluid-fluid demixing phenomena \cite{BinaryExp}. Moreover, in
the presence of external fields colloidal mixtures  also show a
richer behavior than that of monodisperse systems, e.g. oppositely
charged colloids can exhibit lane formation in an electric field
\cite{laneformation,laneformation2,laneformation3}, while
sedimentation in the Earth's gravity field can generate layers of
colloids with equal mass-per-charge \cite{zwanikken} or a
``colloidal Brazil nut effect'' in which a layer of heavier
colloids floats on top of a layer of lighter ones \cite{Hartmut}.
Given the success of the cell model and of the concept of charge
renormalization for the calculation of thermodynamic properties of
monodisperse highly charged colloidal suspensions, as well as the
recent interest and advances in the synthesization of colloidal
mixtures, it is desirable to extend the cell model to the
polydisperse domain. The aim of this paper is to introduce such a
generalized cell model and to study some of its features, in
particular the dependence on the composition of the mixture ---
this plays of course no role in monodisperse suspensions.

The outline of this paper is as follows. In section 2 we introduce
the model. In section 3 we discuss the particular case of a binary
mixture. We calculate saturation curves for the renormalized
charges, osmotic properties, and the radii of the cells as a
function of composition, screening length and total colloid
density. Discussion and some general remarks are presented in
section 4, followed by conclusions in section 5.

\section{Model}

Let us begin by considering two species of spherical colloids with
charges $Q_a=Z_a e$, $Q_b=Z_b e$ and radius $a$ and $b$,
respectively, where $e$ is the proton charge and where we assume
that the colloids possess charges of the same sign, i.e., $Z_a
Z_b>0$. Let $N_a$ and $N_b$ denote the total number of colloids of
type $a$ and those of type $b$ respectively. The colloidal mixture
occupies a volume $V$ and the colloids are immersed in a medium
that is in osmotic equilibrium with a salt reservoir with
screening length $\kappa^{-1}=(8\pi \rho_s \lambda_{B})^{-1/2}$
where $2\rho_s$ is the salt concentration in the reservoir. As
usual we have defined the Bjerrum length $\lambda_B=\beta
e^2/\varepsilon$, with $\varepsilon$ the dielectric constant of
the solvent and $\beta=1/(k_B T)$, with $T$ the temperature and
$k_B$ the Boltzmann constant.  The average number densities of
colloids are defined by $\rho_a=N_a/V$ and $\rho_b=N_b/V$ and
their respective volume fractions are $\eta_a=4 \pi a^{3}
\rho_a/3$ and $\eta_b=4\pi b^3 \rho_b/3$. The composition of the
mixture is characterized by the molar fraction $x=N_a/(N_a+N_b)$
which in the case of equisized colloids can be written as
$x=\eta_a/\eta$ with $\eta=\eta_a+\eta_b$ the total packing
fraction of the suspension.

Following the usual cell model approximation we consider the
Wigner-Seitz cell of each colloidal particle as a sphere concentric
with the spherical colloid. However, due to the difference in size and
charge of each species, there is no a priori reason to assume that
the cells corresponding to each species are of equal size. Therefore, we
suppose that the Wigner-Seitz cells of the colloids of type $a$ have a
radius $R_a$ and those of the colloids of type $b$ have radius $R_b$,
which are in principle different and to this point unknown.

Let $\phi_a(r)=\beta e \psi_a(r)$ be the reduced electric
potential inside a cell of type $a$, with $\psi_a(r)$ the electric
potential at a radial distance $r$ from the center of the colloid.
Accordingly, we define $\phi_b(r)=\beta e\psi_b(r)$ the reduced
electric potential inside a cell of type $b$. In the mean-field
approximation, the electric potential inside each cell is the
solution of the Poisson--Boltzmann equation
\begin{subequations}
  \label{eq:PB1}
\begin{eqnarray}
  \Delta \phi_a(r) &=& \kappa^2 \sinh \phi_a(r)\\
  \Delta \phi_b(r) &=& \kappa^2 \sinh \phi_b(r)
\end{eqnarray}
\end{subequations}
together with the boundary conditions
\begin{eqnarray}
\left.  r\frac{d\phi_a}{dr} \right|_{r=a}=-Z_a \lambda_B /a
\,;
& &
\left.  r\frac{d\phi_b}{dr} \right|_{r=b}=-Z_b \lambda_B /b
\label{eq:Gauss}
\\
\left.  \frac{d\phi_a}{dr} \right|_{r=R_a}=0 \,; & & \left.
\frac{d\phi_b}{dr} \right|_{r=R_b}=0, \label{eq:Gauss-neutral}
\end{eqnarray}
which are obtained from direct application of Gauss' law at the colloid
surface (Eq.~(\ref{eq:Gauss})), and at the boundary of the cell
(Eq.~(\ref{eq:Gauss-neutral})). The latter equation accounts for the
electroneutrality of each cell. These equations completely determine
the electrostatic potential inside each cell. However, the radius of
each cell $R_a$ and $R_b$ remain undetermined, so far. These are
determined by the following relations:
\begin{equation}
  \label{eq:cell-volume}
  \eta_a \frac{R_a^3}{a^3} +
  \eta_b \frac{R_b^3}{b^3} = 1
  \,,
\end{equation}
which expresses the fact that the Wigner-Seitz cells fill the whole
accessible volume $V$, and
\begin{equation}
  \label{eq:phi-continuity}
  \phi_{a}(R_a)=\phi_{b}(R_b)
  \,,
\end{equation}
which is a consequence of the continuity of the electric
potential, and the fact that cells of different species can be in
contact with each other. The system of differential and algebraic
equations~(\ref{eq:PB1}), (\ref{eq:Gauss}),
(\ref{eq:Gauss-neutral}), (\ref{eq:cell-volume})
and~(\ref{eq:phi-continuity}) is a complete set to determine the
unknown parameters and functions $R_a$, $R_b$, $\phi_a(r)$ and
$\phi_b(r)$. The numerical solution of this system of equations
can be obtained with a generalization of the elegant algorithm for
a monodisperse colloidal suspension proposed in
Ref.~\cite{Trizac}.

The generalization to a mixture of $M>2$ colloidal species with
radii $r_i$, packing fractions $\eta_i$ and charges $Z_i$, is
performed by introducing $M$ cells of radii $R_i$
 ($i=1,2,...,M$), in which the
Poisson--Boltzmann equation~(\ref{eq:PB1}) with boundary
conditions~(\ref{eq:Gauss}) and~(\ref{eq:Gauss-neutral}) is to be solved. To fix the
values of the $M$ radii of the cells, the equivalent of
Eq.~(\ref{eq:cell-volume}) is
\begin{equation}
  \sum_{i=1}^M \eta_i \frac{R_i^3}{r_i^3}=1.
\end{equation}
Additionally, we have $M-1$ independent equations of continuity for the electric potential at
the cell boundaries of the form
\begin{equation}
  \phi_i(R_i)=\phi_{1}(R_{1})\,.
\end{equation}
for $i=2,...,M$. This completes the system of equations for the
electric potentials and the radii of the cells.

\subsection{Renormalized charge}

The renormalized charge, defined using Alexander
prescription~\cite{Alexander}, is obtained by comparing the
nonlinear solution of the Poisson--Boltzmann equation with its linearized solution, where the
linearization is performed with respect to the
value of the potential at the boundary of the cell. Following
Ref.~\cite{Trizac}, and restricting the analysis to a binary mixture for the sake of simplicity, let us define $\tilde{\phi}_{i}(r)=\phi_{i}(r)
- \phi_0$, for each cell $i=a,b$, with
$\phi_0=\phi_{a}(R_a)=\phi_{b}(R_b)$. Supposing that
$\tilde\phi_i$ is small, it satisfies the linearized
Poisson--Boltzmann equation
\begin{equation}
\label{eq:PB}
  \Delta \tilde {\phi}_{i}(r)=\kappa_{\PB}^2 (\tilde{\phi}_{i}(r)+\gamma_0)
\end{equation}
with $\kappa_{\PB}^2=\kappa^2 \cosh\phi_0$ and
$\gamma_0=\tanh\phi_0$, and the boundary conditions
$\tilde\phi_{i}(R_{i})=0$ and $\tilde{\phi}_{i}'(R_{i})=0$. The
solution of the linearized problem is
\begin{eqnarray}
  \tilde{\phi}_{i}(r) &=&\gamma_0 \Big[-1
  + \frac{ e^{\kappa_{\PB}
  (r-R_{i})}(\kappa_{\PB}R_{i}+1)}{2\kappa_{\PB} r}
  \nonumber \\
  &&
  + \frac{ e^{\kappa_{\PB}
  (R_{i}-r)}(\kappa_{\PB} R_{i}-1)}{2\kappa_{\PB} r}
  \Big]
  \,.
\end{eqnarray}
The linear potential $\tilde{\phi}_{i}(r)+\phi_0$, the behavior of
which approximates the nonlinear solution $\phi_{i}(r)$ near the
cell boundary, is created by an effective or renormalized charge
$Z_{i}^{\ren}=r_{i}^{2} \tilde{\phi}_{i}'(r_i)/\lambda_{B}$ where
$r_{i}$ is the colloid radius ($r_a=a$, $r_{b}=b$ for $i=a, b$
respectively). Explicitly,
\begin{eqnarray}
  \label{eq:Zren}
  Z_{i}^{\ren} \frac{\lambda_B}{r_i} & = &
  \frac{\gamma_0}{\kappa_{\PB} r_i}
  \Big[
    (\kappa_{\PB}^{2}R_{i} r_{i} -1)
    \sinh(\kappa_{\PB}(R_{i}-r_{i}))
    \nonumber
    \\
    &&
    +\kappa_{\PB} (R_{i}-r_{i}) \cosh(\kappa_{\PB}(R_{i}-r_{i}))
    \Big]\,.
\end{eqnarray}

As mentioned before, a convenient numerical algorithm to compute
the renormalized charges can be formulated on the basis of the
method introduced in Ref. \cite{Trizac}: we impose a value
$\phi_0$ for common value of the electric potential at the cell
boundaries. Solving numerically the system of
equations~(\ref{eq:PB1})--(\ref{eq:phi-continuity}) allows us to
determine the radius of the cells $R_a$ and $R_b$ and the bare
charges $Z_a$ and $Z_b$ corresponding to the value $\phi_0$ of the
potential at the boundaries of the cells. Finally, using
Eq.~(\ref{eq:Zren}) we compute the corresponding renormalized
charges $Z_a^{\ren}$ and $Z_b^{\ren}$. The osmotic properties of
the suspension can also be determined by this procedure as
discussed below.

\subsection{Thermodynamic properties}

In the mean field approximation, the grand potential per cell
($\omega_{a}$ or $\omega_{b}$) is given by (see e.g. Ref \cite{GroumbergCell})
\begin{eqnarray}
  \beta \omega_{i} &=& \frac{1}{8\pi \lambda_B}
  \int \left|\nabla \phi_{i}(r)\right|^2\,d\r
  \nonumber\\
  && + \int \sum_{\alpha=\pm}\left[\rho_{i}^{\alpha}(r) (\ln
  \frac{{e^{\beta\mu_0}\rho_{i}^\alpha}(r)}{\rho_s}-1)\right]\,d\r
  \nonumber\\ &&
    -\int{\beta\mu_0[\rho_i^{+}(r)+\rho_i^-(r)-2\rho_s]}d\r,
 \end{eqnarray}
where the ionic density profiles inside each cell are given by
\begin{eqnarray}
  \rho_{i}^{\pm}(r)= \rho_s e^{\mp \phi_{i}(r)}.
\end{eqnarray}
The integration is performed in the cells excluding the volume
occupied by the colloid, i.e., $r_i<r<R_i$.

Using Gauss' law, Eqs~(\ref{eq:Gauss}) and
(\ref{eq:Gauss-neutral}), this can be written as
\begin{eqnarray}
\beta \omega _{i}&=&\frac{Z_{i}}{2}\phi_{i}(r_i)
\\
&+&4\pi \rho_s \int_{r_{i}}^{R_{i}} dr\, r^{2}\left[
\phi_{i}(r)\sinh \phi_{i}(r) -2\cosh \phi_{i}(r)\right] \,.
\nonumber
\end{eqnarray}
The total grand potential of the system is
\begin{equation}
   \Omega= N_{a} \omega_{a} + N_{b} \omega_{b}+\textsl{F}^{id}_{a}+\textsl{F}^{id}_{b}
\end{equation}
where $\beta\textsl{F}^{id}_{i}={N}_i[\ln({\rho}_i\Lambda_i^3)-1]$
accounts for the colloid entropy, with $\Lambda_i$ and $N_i$ the
thermal length of colloids of species $i$ and their number
respectively. The electrostatic contribution to the pressure $p$
can be obtained deriving $\Omega$ with respect to the volume $V$.
Introducing the osmotic pressure $\Pi=p-p_{res}$ with $p_{res}$
the pressure in the electrolyte reservoir, the final result can be
expressed in terms of the potential at the boundary of the cells
$\phi_0$ in the form
\begin{equation}
\label{CompressFac} \frac{\beta\Pi}{\rho}=
1+\frac{2\rho_s}{\rho}(\cosh \phi_0-1)
\end{equation}
with $\rho=\rho_a+\rho_b$ the total number density of colloids.
The l.h.s. of Eq (\ref{CompressFac}) defines the osmotic
compressibility factor which equals one in the dilute limit. In
the next section we apply the model formulated above to study
charge renormalization and osmotic properties in a colloidal
binary mixture.

\section{Colloidal binary mixture}

Let us consider a colloidal mixture consisting of two species of
colloids. In order to reduce the number of parameters, we will
assume that the colloids have the same radius, i.e., $a=b$ and we
will express all lengths in units of this quantity. In addition,
the Bjerrum length will be fixed to $\lambda_B/a=0.0022$ which is
a typical value for a suspension of colloidal particles in water.
The remaining free parameters are (i) the bare charge of species
$a$: $Z_a$, (ii) the bare charge of species $b$: $Z_b$ (iii) the
total packing fraction: $\eta$, (iv) the composition of the
mixture: $x$, and (v) the concentration of electrolyte as given by
the parameter $\kappa{a}$, where $\kappa^{-1}$ is the screening
length of the electrolyte reservoir. Notice that the screening
parameter $\kappa$ is \emph{different} from $\kappa_{PB}$ as
defined after equation (\ref{eq:PB}), the later corresponding to
the electrolyte concentration \emph{at the cell boundary}.

\subsection{Renormalized charges}

Let us start our analysis by considering the behavior of the
renormalized charges of the binary mixture as a function of the
structural charges. Fig.~1 shows the renormalized charge for each
species when the bare charge of one of the species is fixed while
that of the other species is changed. In Fig.~1a the structural
charge of species $b$ is fixed to $Z_b\lambda_B/a=5$ while $Z_a$
varies. In Fig.~1b we fixed $Z_a\lambda_B/b=10$ and vary the bare
charge of species $b$. Two different compositions are considered,
$x=0.05$ and $x=0.95$, and two screening lengths $\kappa{a}=0.1$
and $\kappa{a}=1$. The remaining parameters are fixed to
$\eta=0.1$, $\lambda_B=0.72$nm and $a=b=326$nm. In Fig.~2 a
similar study is performed, this time for a larger total packing
fraction of the mixture $\eta=0.3$. A first observation following
Figs.~1 and~2 is that the dependence of $Z^{\ren}_i$ is similar to
the behavior found in monodisperse systems: as the bare charges
$Z_i$ increases, the corresponding renormalized charges
$Z_i^{ren}$ grows. The growth is only linear for small values of
the varying structural charges. For larger values of
$Z_i\lambda_B/a$ the increment becomes highly non-linear (compared
to the dashed lines corresponding to $Z^{ren}_i=Z_i$) and
saturates towards a value much smaller than the structural charge
for $Z_i \lambda_B/a \gg 1$. Notice that the saturation value of
the charges depends not only on the total packing fraction but
also on the composition of the mixture. The influence of the
electrolyte concentration on the saturation curves, on the other
hand, is relatively small, as compared to the effect produced by a
drastic change in composition or total packing fraction of the
mixture. In particular, in Fig.~2 the curves corresponding to
$\kappa{a}=1$ and $\kappa{a}=0.1$ are indistinguishable. For the
packing fractions considered in these examples, reducing the
electrolyte concentration manifest itself in a slight reduction of
the saturation values in all cases.

\begin{figure}[tbp]
\begin{center}
\includegraphics[width=70mm,angle=-90]{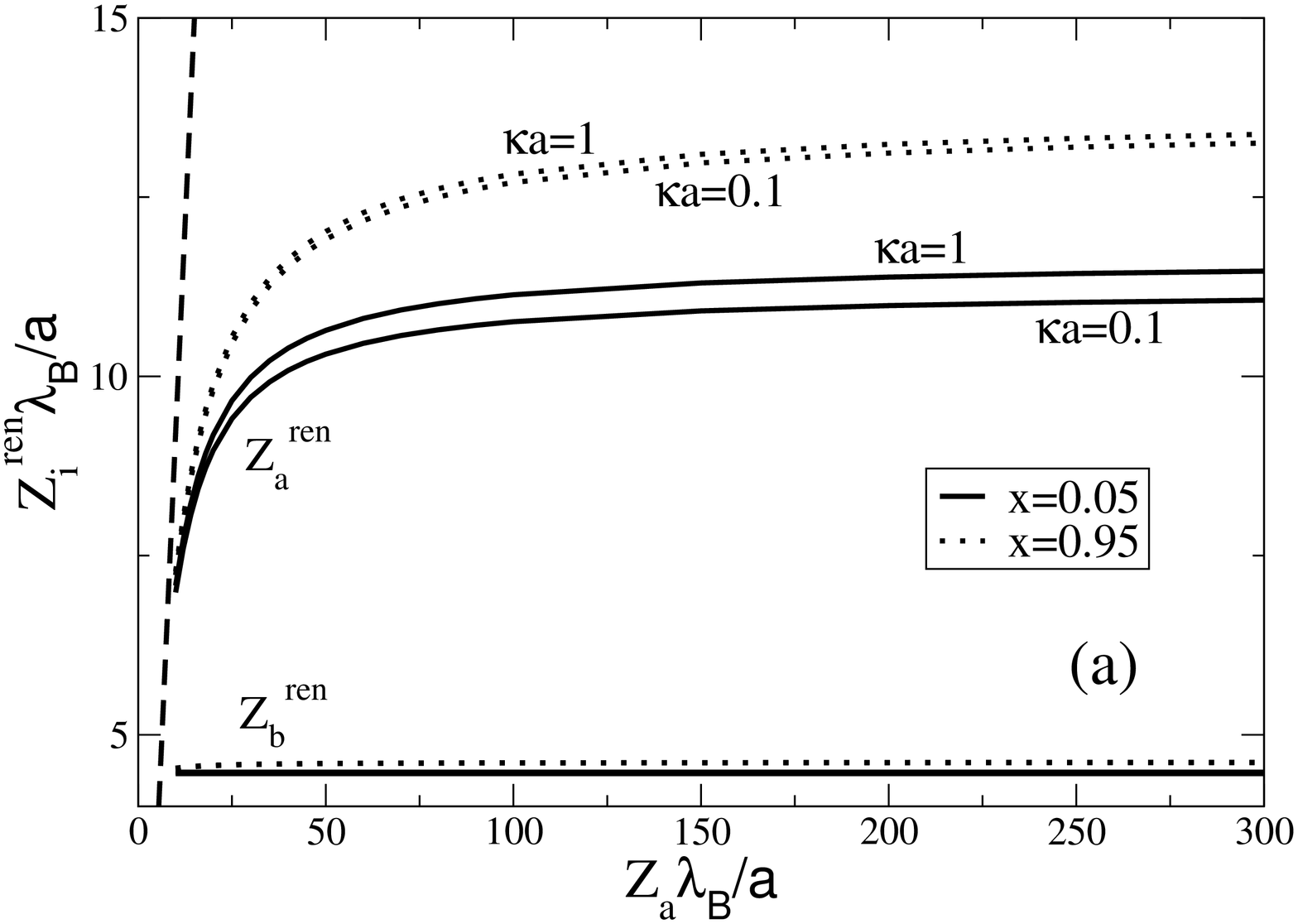}
\hfill
\includegraphics[width=70mm,angle=-90]{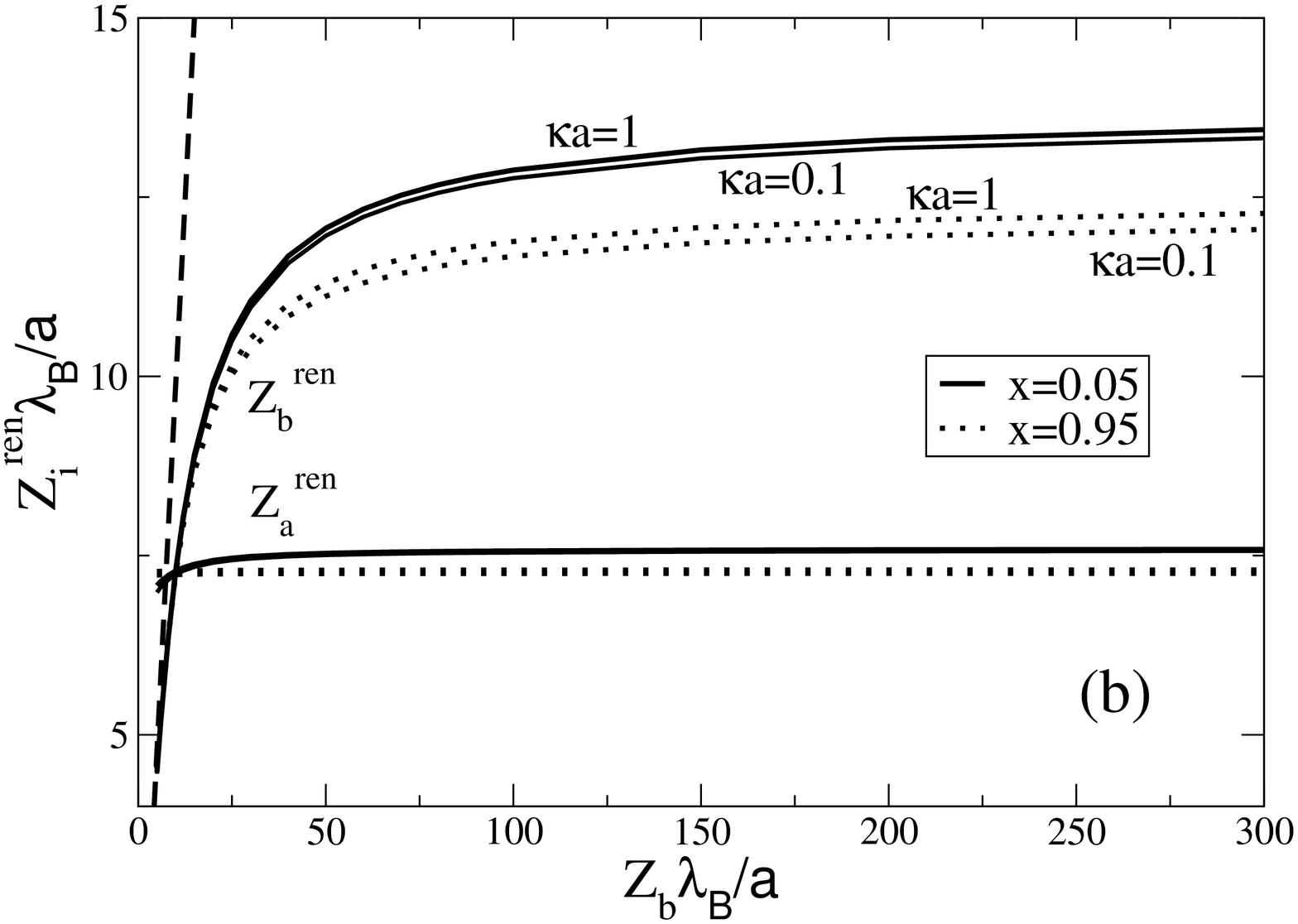}
\end{center}
\caption{Renormalized charges $Z_a^{ren}\lambda_B/a$ and
$Z_b^{ren}\lambda_B/a$ for a binary mixture with: (a) fixed bare
charge $Z_b\lambda_B/a=5$ and varying bare charge $Z_a\lambda_B/a$
and (b) fixed bare charge $Z_a\lambda_B/a=10$ and varying bare
charge $Z_b\lambda_B/a$ for $\lambda_B=0.72$nm and $a=b=326$nm.
The total packing fraction is fixed to the value $\eta=0.1$.
Different electrolyte concentrations and compositions are
considered as pointed by the legends. For the sake of comparizon,
the dashed line shows the linear dependence corresponding to the
case $Z^{ren}_i=Z_i$.}
\end{figure}

\begin{figure}[tbp]
\begin{center}
\includegraphics[width=70mm,angle=-90]{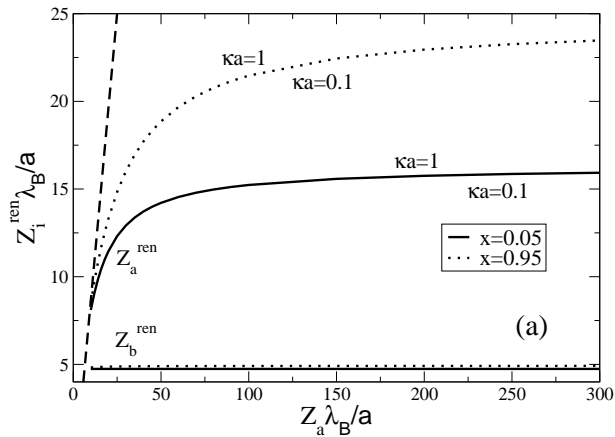}
\hfill
\includegraphics[width=70mm,angle=-90]{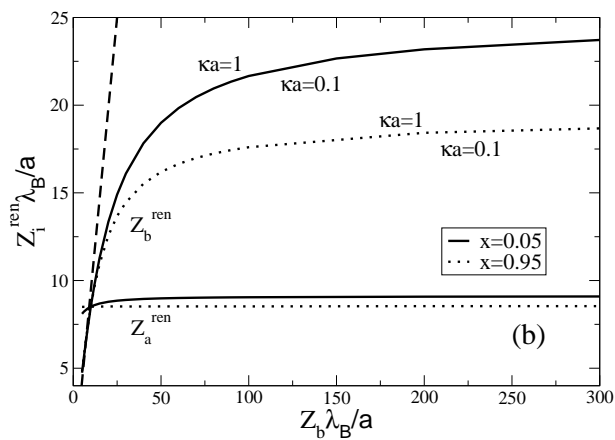}
\end{center}
\caption{Renormalized charges for a binary mixture as given in
Fig.~1. Here we consider a higher value of the total packing
fraction, $\eta=0.3$.}
\end{figure}

The concentration of colloidal particles in suspensions can be
adjusted and measured in a wide range of packing fractions,
therefore, it is
 interesting to analyze to what extent the general behavior of the renormalized
 charges described above depends on this quantity. In Fig.~3 we consider the scaled
 renormalized charges corresponding to a total packing fraction varying over several
 decades. The bare charges of species $a$ and $b$ are fixed by $Z_a\lambda_B/a=10$ and
 $Z_b\lambda_B/a=5$ respectively. The figure shows curves obtained for different electrolyte concentrations as fixed
 by the parameter $\kappa{a}$. Several values of $x=\eta_a/\eta$ are represented by different types of lines.
For $\kappa{a}\gtrsim{1}$ and $\eta\lesssim{0.01}$ varying either
the electrolyte concentration, the total packing fraction, or the
composition of the mixture has a small effect on the values of the
renormalized charges. For $\eta\gtrsim{0.01}$ varying the
composition of the mixture causes the charge renormalization
curves to split into several branches. This effect is also present
for smaller total packing fractions in the extremely low
electrolyte concentration curve corresponding to $\kappa{a}=0.1$.
In the latter case, the values of the renormalized charges are
much more sensitive to variations of the total packing fraction,
which increases after reaching a minimum at about $\eta=0.015$. At
total packing fractions $\eta\gtrsim0.2$ all the curves show the
same qualitative behavior and tend to merge into lines of similar
slope.

\begin{figure}[tbp]
\begin{center}
\includegraphics[width=70mm,angle=-90]{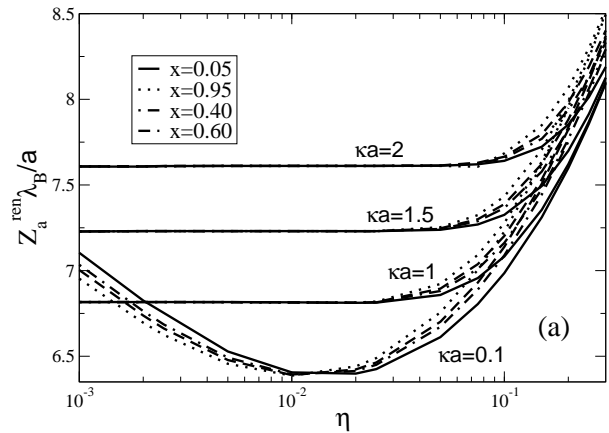}
\hfill
\includegraphics[width=70mm,angle=-90]{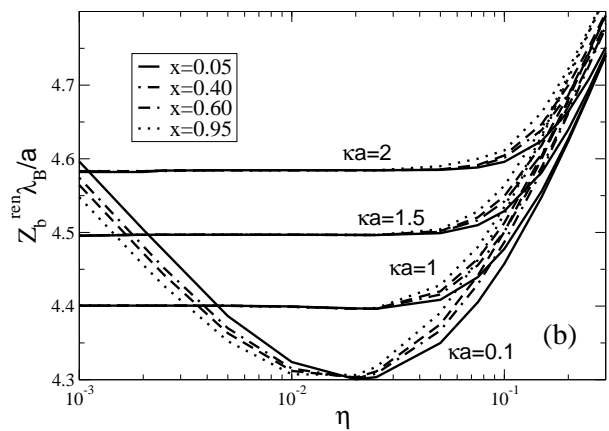}
\end{center}
\caption{Renormalized charges $Z_a^{ren}\lambda_B/a$ and
$Z_b^{ren}\lambda_B/a$ for a binary mixture with bare charges
$Z_a\lambda_B/a=10$ and $Z_b\lambda_B/a=5$ as a function of the
total packing fraction $\eta$. Different electrolyte
concentrations as fixed by the parameter $\kappa{a}$ are shown.
The values of the remaining parameters are typical for colloidal
suspensions, namely $\lambda_B=0.72$nm and $a=b=326$nm. Different
types of lines correspond to distinct compositions of the mixture
as indicated by the legends.}
\end{figure}

\begin{figure}[tbp]
\begin{center}
\includegraphics[width=70mm,angle=-90]{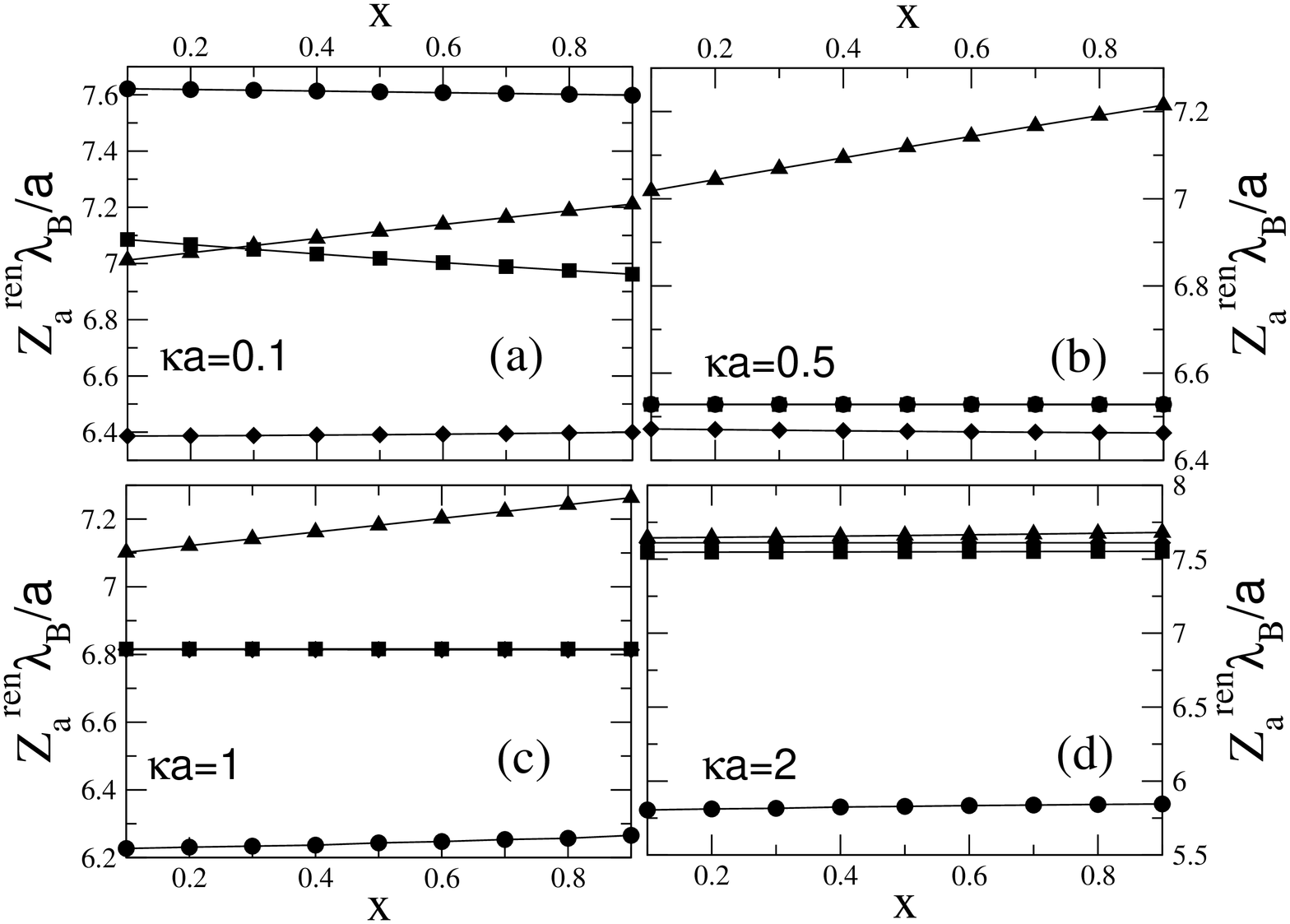}
\end{center}
\caption{\label{fig:Zren-xa} Renormalized charge $Z_a^{ren}
\lambda_B/a$ as a function of the composition $x=\eta_a/\eta$ for
different salt concentrations for the parameters $b=a$, $Z_a
\lambda_B/a=10$, and $Z_b \lambda_B/a=5$.  We consider different
values of the total packing fraction, namely $\eta=10^{-4}$,
(circles), $\eta=10^{-3}$, (squares), $\eta=10^{-2}$ (diamonds),
and $\eta=10^{-1}$ (triangles).}
\end{figure}

In Fig.~4 we show the effect of varying the composition of the
mixture on the scaled renormalized charge of species $a$. We
observe that $Z_a^{ren}\lambda_B/a$ varies linearly with changing
$x$ between the corresponding values of a pure species$-a$ system,
i.e., $x=1$ and a pure species$-b$ system, i.e., $x=0$. Similar
curves are obtained for $Z_b^{ren}\lambda_B/a$. The slope of the
renormalized charge curves is slightly larger for small
electrolyte concentration and is practically zero for $\kappa
a=2$. For the larger packing fractions (triangles), the general
tendency is that the renormalized charges increase with $x$. For
smaller packing fractions and low electrolyte concentration
(squares in Fig.~4a), this tendency is reversed in correspondence
with the behavior observed in Fig.~3a.

\subsection{Osmotic properties}

In Fig.~5 we show the compressibility factor (Eq.
(\ref{CompressFac})) as a function of the total packing fraction,
corresponding to the four different electrolyte concentrations
considered in Fig~3. Different types of lines illustrate the
effect of varying the compositions of the mixture. In general,
adding particles of species with the larger charge increases the
compressibility factor. This effect is particularly notorious at
high packing fractions. At low packing fractions the electrostatic
pressure reduces to the value of the reservoir pressure and the
compressibility factor decays towards the ideal gas value, i.e.,
$\beta\Pi/\rho=1$. It is interesting to observe the strong effect
of varying the electrolyte concentration: following the sequence
from Fig.~5a to Fig.~5d one notices that the compressibility
factor decreases by about two orders of magnitude for $\eta<0.1$.
This diminishing in the compressibility factor is a consequence of
the decreased range of the repulsions. Moreover, for low
electrolyte concentrations the compressibility factor rapidly
grows as a function of the total packing fraction. This effect is
intensified by increasing the concentration of the highest charged
species, species $a$ in this particular case study. For higher
electrolyte concentrations, Fig.~5d, the rate of change of the
compressibility factor is reduced and it assumes smaller values as
comparison with Fig.~5a shows, i.e., the system becomes ideal-like
in the dilute regime, as a consequence of the screening of the
strong repulsive electrostatic forces. This can be seen more
clearly from the osmotic equation of state shown in Fig.~6 for
several electrolyte concentrations. In Fig.~6a, corresponding to
extremely low screening, the osmotic pressure assumes large values
and the suspension is expected to be in crystal phase. After
increasing the screening constant, the osmotic pressure assumes
values close to zero which means that the excess pressure due to
electrostatic interactions reduces to the reference pressure in
the reservoir.

\begin{figure}[tbp]
\begin{center}
\includegraphics[width=70mm,angle=-90]{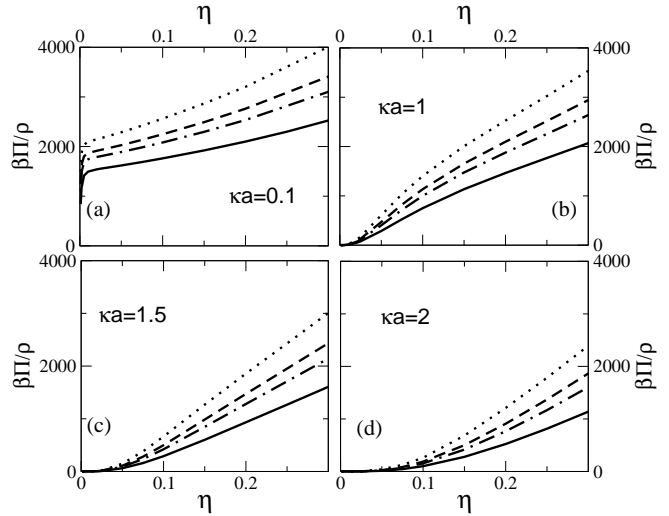}
\end{center}
\caption{Compressibility factor as obtained from Eq.
(\ref{CompressFac}) as a function of the total packing fraction
for different electrolyte concentrations. The line-type convention
for different compositions the same as in Fig.~3.}
\end{figure}

\begin{figure}[tbp]
\begin{center}
\includegraphics[width=70mm,angle=-90]{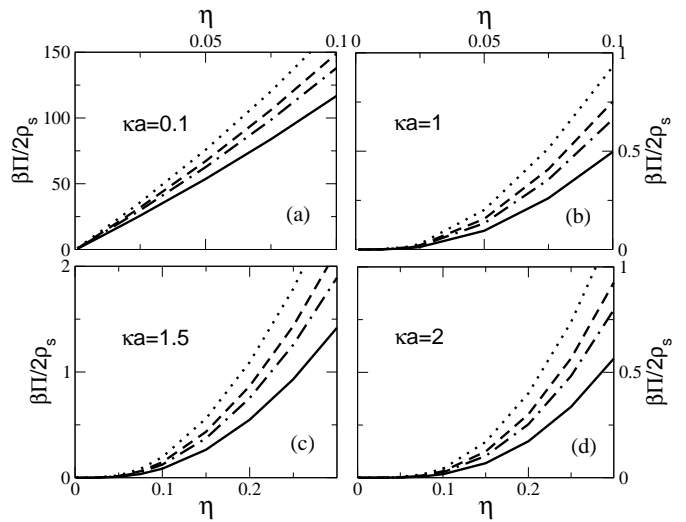}
\end{center}
\caption{Osmotic pressure as a function of the total packing fraction
for different electrolyte concentrations. The line-type convention
is the same as in Fig.~3.}
\end{figure}

\subsection{Cell radii}

\begin{figure}[tbp]
\begin{center}
\includegraphics[width=70mm,angle=-90]{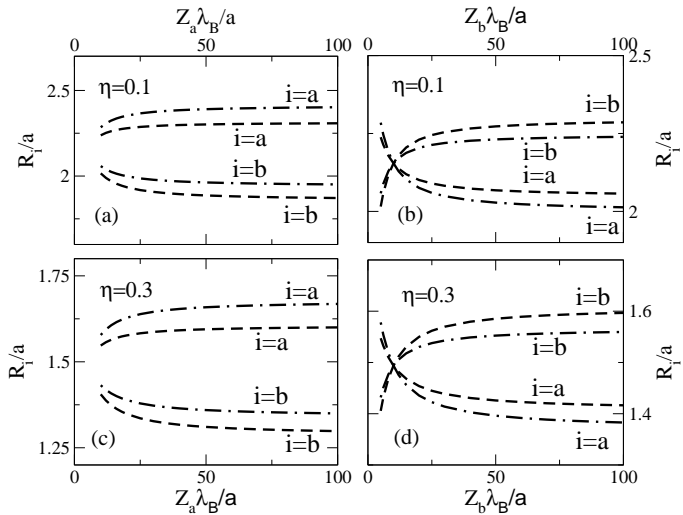}
\end{center}
\caption{Radii of the cells $R_a$ and $R_b$ surrounding particles
of type $a$ and $b$ respectively expressed in units of the
colloids radius $a=b=326$nm. In (a) and (c) the bare charge of
type-a particles varies while $Z_b\lambda_B/a=5$ is kept fixed. In
(b) and (c) the bare charge of type-b particles varies while
$Z_a\lambda_B/a=10$ remains unchanged. Two different total packing
fractions are considered as pointed by the legends and the values
$\lambda_B=0.72$nm and $\kappa{a}=1$ are used in all figures. The
line-type convention is the same as that in Fig.~3.}
\end{figure}

In Fig.~7 we plot the cell radii (expressed in units of the
colloidal radius $a$) as a function of the charge of one of the
species while the charge of the other species is fixed (see
details in the caption). In general, as one of the radii increases
the other radius decreases to satisfy the constraint of fixed
total packing fraction  as expected from Eq.
(\ref{eq:cell-volume}). The form in which this occurs depends on
the colloid charges and compositions as well as on the other
physicochemical parameters in the model. A small total packing
fraction allows for larger cell radii as we notice by comparing
Figs.~7a and 7c on the one hand and Figs.~7b and 7d on the other
hand. Notice also that in the latter case for small values of
$Z_b\lambda_B/a$ we have $R_a/R_b>1$. After the charge of species
$b$ is increased, the inequality quickly inverts i.e., the volume
of the cells surrounding the now highly charged particles of
species $b$ surpasses that of species $a$. Finally, it is worth to
notice the fact that beyond some value of $Z_i\lambda_B/a$ the
radii of the cells become almost charge independent, reflecting
the fact that the renormalized charges have reached their
saturation value. In the light of the previous case study, we now
proceed to discuss in more detail some general features of the
polydisperse cell model.

\section{Discussion}

Physically, the charge renormalization process results from the
accumulation of ions in the vicinity of a highly charged object
immersed in an electrolyte. Therefore, the relevant parameter to
compute their interactions is not the bare charge but the
effective (renormalized) charge that takes into account this
non-linear screening effect. As shown in Fig.~1, the behavior of
the renormalized charges for polydisperse systems at the level of
PB theory is reminiscent  of that of their monodisperse
counterpart: the renormalized charges saturate to a finite value
when the structural charge becomes sufficiently large. The
saturation values can be orders of magnitude smaller than the
structural charges. As a consequence of this non-linear screening
effect, a very asymmetric colloidal mixture behaves as a much more
symmetric system as regards their electrostatic effects. For
instance, from Fig.~2b one can notice that a colloidal binary
mixture with charges $Z_a\lambda_B/a=10$ and $Z_b\lambda_B/a=300$
behaves \emph{effectively} as a mixture with
$Z_a^{ren}\lambda_B/a\simeq7.5$ and
$Z_b^{ren}\lambda_B/a\simeq13$, so that the ratio $Z_a/Z_b=0.03$
changes to $Z_a^{ren}/Z_b^{ren}=0.58$ which corresponds to a much
more charge symmetric effective mixture. On the other hand, for
small values of the structural charges, $Z^{ren}_i\simeq{Z}_i$. As
a consequence, expressions for the effective interaction
potentials derived from linear theories, e.g., DLVO theory, are
expected to be sufficient to describe the thermodynamic properties
of the suspension, at least in the regime where pairwise
additivity is expected \cite{BelloniSed,torres_pre}.

The saturation values of the renormalized charges depend on the
physicochemical parameters: the electrolyte concentration, total
colloids density, the dielectric constant of the solvent and the
composition of the mixture. Increasing the electrolyte
concentration has the effect of enhancing the value of the
renormalized charges. This effect induces two competing tendencies
in the screened interaction potential. On one hand the spatial
extent of the double-layer reduces, on the other hand the
amplitude of the interactions increases. This tendency is inverted
in the case of weak screening and total colloids packing fraction
smaller that about $5\%$ (see Fig.~3). The balance of these two
non-linear screening features is affected by the composition of
the mixture and is expected to have consequences on, e.g., the
phase diagrams of mixtures of highly charged macroions computed by
using Yukawa-like potentials, the omission of such non-linear
screening effects, as accounted by $Z_i^{ren}$, may lead to
unphysical results as demonstrated by several authors (See e.g.
\cite{Levin,Levin2}).

 In the context of the cell model for monodisperse systems, it is
 assumed that due to entropic reasons, the colloidal particles tend to
 occupy the maximum space possible inside the solvent. Therefore, it
 is feasible to factorize the suspension into spherical cells, each of
 them housing a colloidal particle in its center.  The radius of such
 cells is then chosen in such a fashion that the cells fill all the
 suspension, namely $R/a=\eta^{-1/3}$.  In other words, the packing
 fraction alone determines the size of the cells. In contrast, in the
 polydispense cell model, the radii of the cells depend not only on
 the total packing fraction and the composition of the mixture
 according to Eq. (\ref{eq:cell-volume}), but also on the relative
 charges of the species through the boundary conditions expressed by
 Eqs. (\ref{eq:Gauss-neutral}) and (\ref{eq:phi-continuity}). In other
 words, the volumes of the cells are not only determined by fixing the
 total packing fraction, but adapt themselves in order to satisfy the
 boundary conditions which impose the cells to be electrically
 neutral, the potential to be continuous at the cells boundaries and
 the cells to fill the whole volume accessible to the colloidal
 particles. The resulting cells are independent units regarding
 electrostatic interactions in virtue of their neutrality. Therefore,
 the grand potential of the entire suspension can be easily calculated
 as the sum of the grand potential of individual cells. This
 facilitates enormously the calculation of thermodynamic response
 functions, in particular, following the contact theorem, (see e.g.,
 Ref. \cite{Israel}), the osmotic pressure can be obtained directly
 from the value of the electrostatic potential at the cells boundary
 as explained in Sec.~2. The value of the potential at the cell
 boundary depends on the size of the cell, which have a priori
 different radii depending on the parameters, and in particular, on
 the composition of the mixture.

Experimentally, in spite of the recent advances in electrophoretic
techniques \cite{Vladimir}, it is difficult to precisely measure
the values of renormalized charges in colloidal suspensions.
Therefore, it would presumably be intricate to collect
experimental evidence of, e.g.,
 the splitting of the charge renormalization curves in Fig.~3, even at the highest packing fractions.
 Osmotic properties are nevertheless relatively easy to probe experimentally e.g., in sedimentation-diffusion
 equilibrium \cite{Piazza},\cite{Experiments}.
 In that sense the present formalism provides a reasonably simple and numerically
 robust scheme for the calculation of the osmotic equation of state and
 compressibility factors in a wide range of parameters. As illustrated in the case study of Sec.~2, these osmotic properties
 are non-trivially dependent on the mixture composition, which illustrates the
 fact that the relatively small change in the values of the corresponding renormalized charges
 shown in Fig.~3, \emph{do not} necessarily imply a mild change in the response functions: the former
 are obtained from changes in the slope of the potential at the colloids surface as given by the solution of the
 linearized PB equation (see Eq. \ref{eq:Zren}), whereas the latter depend on the value of the potential at the cells
 boundary (Eq. \ref{CompressFac}). The screening parameter, however, plays a much more important role in the osmotic properties
 than the relative composition of the mixture, and may cause, for instance, the osmotic pressure to vary over several orders of magnitude.

It is worth to mention the fact that in the polydisperse cell
model we restricted the attention to colloids of the same sign of
charge. The reason is that one can impose on a physical ground
that each cell surrounding a colloid is charge neutral i.e.,
$\phi_i'(R)=0$ for $i=1,...,M$. It is not obvious that such a
condition can be imposed for oppositely charged colloids, as the
potential may change sign in between the colloids while the
electric field does not vanish anywhere. The attractive or
repulsive nature of the interactions between the cells is not
obvious to estimate by analyzing the sign of the potential, since
e.g., potentials opposite in sign can still lead to colloidal
repulsions \cite{langmuir}.  A second limitation related to the
cells electoneutrality, which is also present in Alexander's
formulation, is that as a consequence of the restriction of the
theoretical analysis to a single cell per colloidal component, all
information on correlations between the colloids is lost. This may
lead to unphysical ideal-gas-like values for the osmotic pressure
in highly screened suspensions \cite{Marcus}. Care has to be
exercised in drawing conclusion from cell models in this regime.

   An important feature of charge renormalization is the fact that increasing the electrolyte
   concentration leads to higher values of the renormalized charges simultaneously reducing
   the extension of the double-layer \cite{Trizac1,Trizac2}. These two non-linear screening features
   are neglected in linear theories, such as the DLVO model, where the (structural) charge is in
   principle independent of the screening parameter. These artifacts of the linearization procedure
   can be cured by using the renormalized values of the charges within the linear schemes, which
 can in principle be achieved for example by using the analytical solution of the Ornstein-Zernike equation
 for a mixture interacting via a DLVO-like potential within Mean Spherical Approximation, by replacing $Z_i\rightarrow{Z_i^{ren}}$.
 A second alternative is provided by the variational treatment of DLVO-like potentials based on the Gibbs-Bogoliubov inequality,
 which uses the Laplace transforms of the pair correlation functions to estimate the free energy of the system
 though a numerical optimization problem (see e.g., Refs. \cite{Boublik, Rosenfeld eqs}). This alternative is
 extremely suitable for the calculation of phase diagrams by e.g., using Maxwell constructions, providing a
 simple alternative for the study of phase behavior of highly charged colloidal mixtures while fully taking into account non-linear effects.

\section{Conclusion}

In this paper we proposed a simple model which allows to study the
thermodynamic properties of mixtures of highly charged macroions
suspended in an electrolyte. The model is a generalization of the
well-know cell model for colloids, and the notion of charge
renormalization as introduced by Alexander and collaborators
\cite{Alexander}. In contrast to the one-component cell model,
where the radius of the cells is fixed by the packing fraction of
colloids, in the polydisperse cell model the radii of the cells
depend on the physicochemical parameters through the boundary
conditions utilized to solve the non-linear Poisson-Boltzmann
equation in the cell geometry. The radii of the cells play the
role of Lagrange multipliers that enforce the constraint that the
cells fill the entire volume occupied by the colloidal mixture and
have to be determined within the numerical solution of the
non-linear differential equations. We presented a method to
perform such a calculation based on the algorithm of Ref.
\cite{Trizac}, which also allows to determine thermodynamic
response functions and renormalized charges within a clear and
efficient numerical scheme. We presented a detailed example of
such an analysis in the case of a colloidal binary mixture, where
we studied the effect that several physicochemical parameters have
on the values of the saturation curves for the renormalized
charges and osmotic properties of the suspension. In general, the
ion condensation phenomenon that leads to charge renormalization
has the effect of making
   the mixture more ``electrostatically symmetric'', in the sense that the ratio of the renormalized charges is much closer to one
   than that of the structural charges. Increasing the concentration of the species with the
   greatest charge keeping the total packing fraction fixed, raises the osmotic pressure. An important limitation of the present model, which is actually shared by all Poisson-Boltzmann-based models,
is the fact that ion-ion correlations are neglected. This oversimplification of the problem may lead not only
to quantitative but also qualitative modifications to the saturation curves: the renormalized charge may exhibit
a maximum followed by a decrease of $Z^{ren}_i$ with $Z_i$ increasing
(see e.g., Ref. \cite{BelloniRev, LegerLevesque,CamargoTellez}).
In some cases this can lead to a weakening
of the repulsive effective interactions as calculated on the basis of the renormalized charges, which may eventually
turn electrostatic repulsions into attractions. In the case of 1:1 electrolytes, such ion-ion correlations are
relatively unimportant and the present Poisson-Boltzmann model is expected to provide a reasonably good description
of highly charged colloidal mixtures.

\begin{acknowledgments}
This work is part of the research program of the "Stichting voor
Fundamenteel Onderzoek der Materie (FOM)," which is financially
supported by the "Nederlandse Organisatie voor Wetenschappelijk
Onderzoek (NWO).
\end{acknowledgments}

\end{document}